\documentclass{PoS}
\usepackage{amsfonts} 
\usepackage{amssymb} 
\usepackage{amsmath} 
\usepackage{graphicx} 
\usepackage{subcaption}
\usepackage{array} 
\usepackage{bm} 
\usepackage{longtable} 
\usepackage{color} 
\usepackage{comment}
\usepackage{verbatim}
\usepackage{multirow}

\graphicspath{{./figs/}}

\allowdisplaybreaks

\newcommand{\MSbar}{{\overline{\text{MS}}}}

\newcommand{\GeV}{\text{GeV}}

\newcommand{\fm}{\,\mathrm{fm}}

\newcommand{\tr}{\textrm{tr}}

\title{Perturbative calculation of $Z_q$ at the one-loop level using
  HYP-smeared staggered quarks}

\ShortTitle{One-loop perturbation for $Z_q$}

\author{\speaker{Benjamin J.~Choi}, Weonjong Lee\\
  Lattice Gauge Theory Research Center, CTP, and FPRD,\\
  Department of Physics and Astronomy,
  Seoul National University, Seoul 08826, South Korea\\
  E-mail: \email{benjaminchoi@snu.ac.kr}, \email{wlee@snu.ac.kr}}

\author{Jangho Kim\\
  Institut f{\"u}r Theoretische Physik, Johann Wolfgang
  Goethe-Universit{\"a}t,\\
  Max-von-Laue-Str.~1, 60438 Frankfurt am Main, Germany\\ 
  E-mail: \email{jkim@th.physik.uni-frankfurt.de}}

\author{Sungwoo Park\\
  Los Alamos National Laboratory, Theoretical
  Division, T-2, Los Alamos, NM 87545, USA\\
  E-mail: \email{kunsung5@gmail.com}}

\author{Stephen R.~Sharpe\\
  Department of Physics, University of
  Washington, Seattle, WA 98195-1560, USA\\ 
  E-mail: \email{srsharpe@uw.edu}}

\author{SWME Collaboration}

\abstract{We present matching factors for $Z_q$ calculated
  perturbatively at the one-loop level with improved staggered quarks.
  We calculate $Z_q$ with HYP-smeared staggered quarks and
  Symanzik-improved gluons using both RI-MOM and RI$'$-MOM schemes.
  We compare the results with those obtained using the nonperturbative
  renormalization (NPR) method.
  }

\FullConference{The 36th Annual International Symposium on Lattice
  Field Theory - LATTICE2018\\ 22-28 July, 2018\\ Michigan State
  University, East Lansing, Michigan, USA.}

\begin{document}

\section{Introduction}
\label{sec:intr}

In order to convert the results for matrix elements from lattice QCD into those in a
continuum scheme, we need to calculate the corresponding renormalization factors.
This can be done either using perturbation theory (PT) or with
a non-perturbative method such as non-perturbative renormalization (NPR). 
The former suffers from truncation errors, while the latter has the usual systematic
errors associated with lattice quantities, as well as the need for a window
in which $\Lambda_{\rm QCD} \ll \mu \ll 1/a$, where $\mu$ is the renormalization scale.
Although NPR is generally preferred, it is useful to make detailed comparisons with PT,
in particular since some lattice calculations of matrix elements use perturbative matching.
Here we present such a comparison for the $Z_q$, the quark field renormalization,
which is an ingredient in NPR calculations for almost all operators.
Specifically, we compare one-loop results for the asqtad action and HYP-smeared
staggered valence quarks 
with those obtained using NPR on the MILC ``coarse" ensemble
($a\approx 0.12\fm$).

\section{Feynman rules}
\label{sec:feyn-self}

%
%
\begin{figure}[t!]
  \begin{subfigure}{0.22\linewidth}
    \vspace*{2.9mm}
    \centering
    \includegraphics[width=\textwidth]{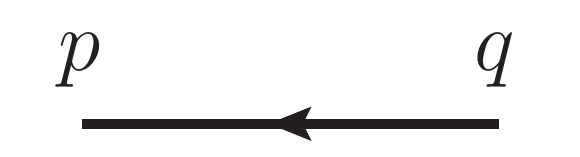}
    \caption{quark propagator}
    \label{fig:q-prop}
  \end{subfigure}
  \hfill
  \begin{subfigure}{0.22\linewidth}
    \vspace*{-7mm}
    \centering
    \includegraphics[width=\textwidth]{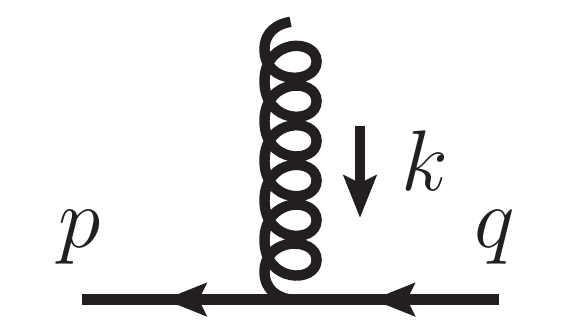}
    \caption{one-gluon vertex}
    \label{fig:1-glu}
  \end{subfigure}
  \hfill
  \begin{subfigure}{0.22\linewidth}
    \vspace*{-7mm}
    \centering
    \includegraphics[width=\textwidth]{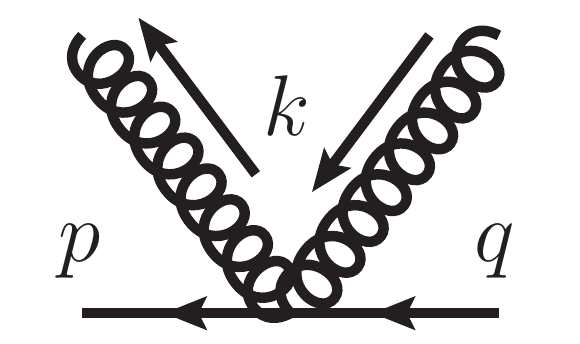}
    \caption{two-gluon vertex}
    \label{fig:2-glu}
  \end{subfigure}
  \hfill
  \begin{subfigure}{0.22\linewidth}
    \vspace*{-5mm}
    \centering
    \includegraphics[width=\textwidth]{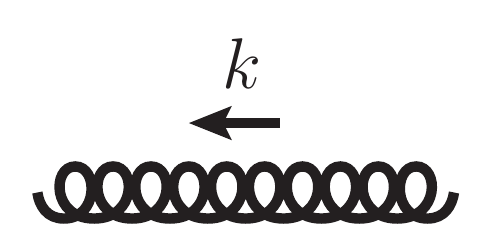}
    \caption{gluon propagator}
    \label{fig:g-prop}
  \end{subfigure}
  \caption{Feynman rules at the one-loop level}
  \label{fig:Feyn-rule}
\end{figure}
The free propagator for HYP staggered quarks in
Fig.~\ref{fig:Feyn-rule}\subref{fig:q-prop} is (with color factors excluded)
\begin{align}
S_{0}\left(p,-q\right) & =\bar{\delta}\left(p'-q'\right) \frac{
  \dfrac{i}{a} \displaystyle \sum_{\mu} \sin \left(a p'_{\mu}\right)
  \overline{\overline{\left(\gamma_{\mu}\otimes\mathbf{1}\right)}}_{AB}
  + m_{0}
  \overline{\overline{\left(\mathbf{1}\otimes\mathbf{1}\right)}}_{AB}
}{\displaystyle \sum_{\alpha} \dfrac{1}{a^{2}} \sin^{2} \left(a
  p'_{\alpha}\right) + m_{0}^{2}}
\end{align}
where $p = p' + \dfrac{\pi}{a}A$, and $q = q' + \dfrac{\pi}{a}B$.
Here, $p'_\mu, q'_\nu \in
\left(-\dfrac{\pi}{2a},+\dfrac{\pi}{2a}\right]$ are momenta
  defined in the reduced Brillouin zone \cite{Patel:1992vu},
$\bar{\delta}\left(p^{\prime}-q^{\prime}\right)$ is the periodic
delta function which is nonzero for 
$q'_{\mu} = p'_{\mu}$ (mod $\dfrac{2\pi}{a}$),
and $A$ and $B$ are hypercubic vectors: $A_\mu, B_\nu \in \{0,1\}$.
The spin-taste factors $\overline{\overline{\left(\gamma_S\otimes \xi_F\right)}}_{AB}$ 
are explained in Ref.~\cite{Patel:1992vu}.

The HYP action uses HYP-smeared links, $V_\mu$, which must be expressed in
terms of the original thin links $U_\mu$. To do so we expand both links as
\begin{equation}
  U_{\mu}(x) = \exp\left[iag
    A_{\mu}\left(x+\frac{a}{2}\hat{\mu}\right) \right] \,,
\qquad
  V_{\mu}(x) = \exp\left[ iag
    B_{\mu}\left(x+\frac{a}{2}\hat{\mu}\right) \right] \,,
\end{equation}
where $A_{\mu}(x)=\displaystyle\sum_{a}A^{a}_{\mu}(x)T^a$ is
the gluon field, while $B_{\mu}(x) = \displaystyle \sum_{a}
B^{a}_{\mu}(x)T^a$ is smeared gauge field.
The latter can be written as a perturbative expansion in
powers of $A_{\mu}(x)$'s
\begin{align}
  B_{\mu}(x) &= \sum^{\infty}_{n=1}B^{(n)}_{\mu}(x) =
  B^{(1)}_{\mu}(x) + B^{(2)}_{\mu}(x) + B^{(3)}_{\mu}(x) + \ldots \,,
\end{align}
where $B_{\mu}^{(n)}$ stands for a term of order $\left( A_{\mu}
\right)^n$.
Only the linear term, 
$B_{\mu}^{(1)}(x)$, contributes to the renormalization at the one-loop
level~\cite{Patel:1992vu,Lee:2002fj}.
The relation between $B_{\mu}^{(1)}(x)$ and $A_{\mu}(x)$ is
\begin{align}
B^{(1)}_{\mu}\left(x\right) &= \int^{\pi/a}_{-\pi/a}\frac{d^4
  k}{\left(2\pi\right)^{4}} \sum_{\nu} h_{\mu\nu}\left(k\right)
\tilde{A}_{\nu}\left(k\right) e^{ik\cdot x}\,, \qquad
A_{\mu}\left(x\right) = \int^{\pi/a}_{-\pi/a}\frac{d^4
  k}{\left(2\pi\right)^{4}} \tilde{A}_{\mu}\left(k\right) e^{ik\cdot
  x} \,,
\end{align}
where $h_{\mu\nu}(k)$ is the smearing kernel which describes details
of the blocking transformation for the fat link.
\begin{align}
h_{\mu\nu}\left(k\right) &=\delta_{\mu\nu} D_{\mu}\left(k\right) +
\left( 1 - \delta_{\mu\nu} \right) \tilde{G}_{\nu,\mu}\left(k\right)
\bar{s}_{\mu} \bar{s}_{\nu}
\\
D_{\mu}\left(k\right) &=1 - d_{1} \sum_{\nu\neq\mu}
\bar{s}_{\nu}^{2} + d_{2} \sum_{\substack{\nu<\rho\\ \nu,\rho\neq\mu}}
\bar{s}_{\nu}^{2} \bar{s}_{\rho}^{2} - d_{3} \bar{s}_{\nu}^{2}
\bar{s}_{\rho}^{2} \bar{s}_{\sigma}^{2} + d_{4} \sum_{\nu\neq\mu}
\bar{s}_{\nu}^{4} 
\\
\tilde{G}_{\nu,\mu}\left(k\right) &=d_{1} - d_{2}
\frac{\bar{s}_{\rho}^{2}+\bar{s}_{\sigma}^{2}}{2} + d_{3}
\frac{\bar{s}_{\rho}^{2}\bar{s}_{\sigma}^{2}}{3} + d_{4}
\bar{s}_{\nu}^{2}\,.
\end{align}
Here, $\mu \neq \nu \neq \rho \neq \sigma$ and
$\bar{s}_{\mu}=\sin(a k_{\mu}/2)$.
In order to remove $\mathcal{O}\left(a^2\right)$ taste symmetry
breaking interactions at tree level, we choose the parameters to be~\cite{Lee:2002ui}
\begin{align}
  d_1 &= 1,\quad d_2=1, \quad d_3=1, \quad d_4=0.
\end{align}

The Feynman rule for the one-gluon emission vertex of
Fig.~\ref{fig:Feyn-rule}\subref{fig:1-glu} is
\begin{align}
V_{\mu;\alpha}^{I}\left(p,-q,k\right) &= - i g T^{I}\cos \left(\frac{a
  k_{\mu}}{2} - a q'_{\mu}\right) h_{\mu\alpha}\left(k\right)
\overline{\overline{\left(\gamma_{\mu}\otimes\mathbf{1}\right)}}_{AB}
\bar{\delta}\left(p'-q'+k\right)\,,
\end{align}
where $T^I$ is the SU(3) color generator,
while that for the two-gluon emission vertex of
Fig.~\ref{fig:Feyn-rule}\subref{fig:2-glu} is
\begin{align}
V_{\mu\nu;\alpha\beta}\left(p,-q,k\right) &= - i a g^{2} C_F
\delta_{\mu\nu}\sin \left(a q'_{\mu}\right)
h_{\mu\alpha}\left(k\right) h_{\mu\beta}\left(k\right)
\overline{\overline{\left(\gamma_{\mu}\otimes\mathbf{1}\right)}}_{AB}
\bar{\delta}\left(p'-q'\right)\,,
\end{align}
where $C_F = \displaystyle \sum_{I} \left(T^I\right)^2={4}/{3}$.

The MILC asqtad ensembles use a Symanzik-improved gluon action.
The corresponding gluon propagator
can be written as~\cite{Kim:2010fj} 
\begin{align}
  \mathcal{D}_{\mu\nu}^{\text{Imp.}}\left(k\right) & = \left(1-\alpha\right)
  \frac{\mathcal{P}_{\mu\nu}}{\hat{k}^{2}} +
  \frac{\left[\hat{k}^{2}\left(\hat{k}^{2}-\tilde{c}x_{1}\right) +
      \tilde{c}^{2}x_{2}\right]\delta_{\mu\nu}^{T} +
    \tilde{c}\left(\hat{k}^{2} -
    \tilde{c}x_{1}\right)\mathcal{M}_{\mu\nu} +
    \tilde{c}^{2}\left[\mathcal{M}^{2}\right]_{\mu\nu}}{f\left\{
    \hat{k}^{2}\left[\hat{k}^{2}\left(\hat{k}^{2} -
      \tilde{c}x_{1}\right) + \tilde{c}^{2}x_{2}\right] -
    \tilde{c}^{3}x_{3}\right\} }\,,
  \label{eq:imp-gluon-prop-1}
\end{align}
where $\alpha = 0 (1)$ for Feynman (Landau) gauge.
The notation in this result is explained in Ref.~\cite{Kim:2010fj}.
Although the gluon action used in generating the MILC lattices includes improvements
beyond tree level, for our one-loop perturbative calculation it is
appropriate to use the tree-level Symanzik improvement coefficients,
with the redundant coefficient chosen according to the 
the convention of Ref.~\cite{Luscher:1984xn}. This choice fixes the constants
in Eq.~\eqref{eq:imp-gluon-prop-1}.
%

\section{Staggered quark self energy}
\label{sec:qse}

%
%
\begin{figure}[t!]
  \begin{subfigure}{0.40\linewidth}
    \vspace*{-7mm}
    \centering
    \includegraphics[width=\textwidth]{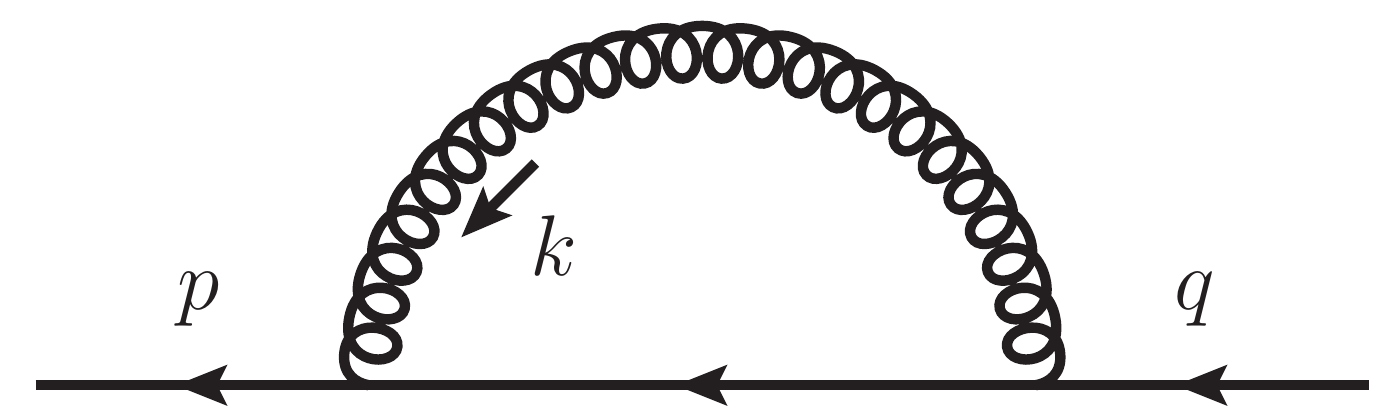}
    \caption{Sunset diagram}
    \label{fig:Z-1}
  \end{subfigure}
  \hfill
  \begin{subfigure}{0.40\linewidth}
    \vspace*{-7mm}
    \centering
    \includegraphics[width=\textwidth]{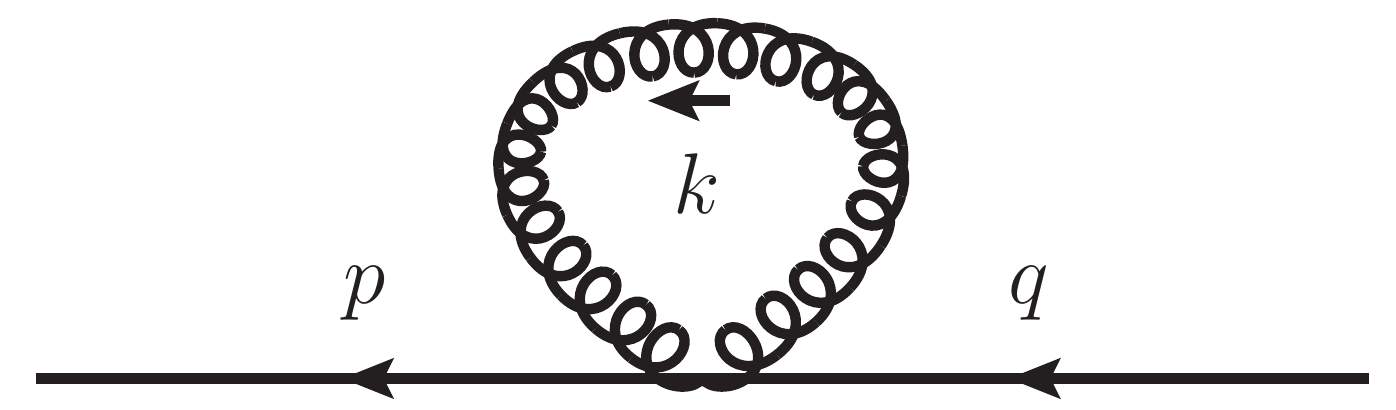}
    \caption{Tadpole diagram}
    \label{fig:ZT-1}
  \end{subfigure}
  \caption{Feynman diagrams for quark self energy at the one-loop level}
  \label{fig:F-diag}
\end{figure}

Using the Feynman rules given above,
we can express the one-loop quark self-energy as
\begin{align}
    \Sigma\left(p,-q\right) &= \int^{\pi/a}_{-\pi/a}\frac{d^4 k d^4 \ell_1
      d^4 \ell_2}{\left(2\pi\right)^{12}}
    \sum_{ \substack{ \mu,\nu, \\ \alpha,\beta} } \sum_{I}
    V_{\mu;\alpha}^{I}\left(p,-\ell_{1},-k\right)
    S_{0}\left(\ell_{1},-\ell_{2}\right)
    V_{\nu;\beta}^{I}\left(-q,\ell_{2},k\right)
    \mathcal{D}_{\alpha\beta}^{\text{Imp.}}\left(k\right)
    \nonumber \\
    &\phantom{=\ } + \frac{1}{2} \int^{\pi/a}_{-\pi/a}\frac{d^4
      k}{\left(2\pi\right)^{4}}
    \sum_{ \substack{ \mu,\nu, \\ \alpha,\beta} }
    V_{\mu\nu;\alpha\beta}\left(p,-q,k\right)
    \mathcal{D}_{\alpha\beta}^{\text{Imp.}}\left(k\right)
    \label{eq:1loop-1}\\
 &= \Sigma\left(p'\right) \bar{\delta}\left(p'-q'\right) \,.
\label{eq:sig1-2-1}
\end{align}
In Eq.~\eqref{eq:1loop-1}, the first term corresponds to
Fig.~\ref{fig:F-diag}\subref{fig:Z-1} and the second to
Fig.~\ref{fig:F-diag}\subref{fig:ZT-1}.

We use the technical tools in Ref.~\cite{Golterman:1984cy} to
calculate the integrals in Eq.~\eqref{eq:1loop-1}.
If we choose the momentum $p'$ so that $|ap'_\mu| \ll 1$,
we can rewrite the self energy as
\begin{align}
\Sigma\left(p'\right) &= -i \Sigma_{1} p_{\mu}^{\prime}
\overline{\overline{\left(\gamma_{\mu}\otimes\mathbf{1}\right)}}_{AB}
+ m_{0} \Sigma_{2}
\overline{\overline{\left(\mathbf{1}\otimes\mathbf{1}\right)}}_{AB}
+ \mathcal O(a^2)
\label{eq:sig1-2-2}
\end{align}
where
\begin{align}
    \Sigma_{1}\left(p'\right) & =
    \frac{g^{2}C_{F}}{\left(4\pi\right)^{2}} \left[
      \left(1-\alpha\right) \left\{ \log\left[ a^{2} \left( m_{0}^{2}
        + p^{\prime2} \right) \right] + \frac{m_{0}^{2}}{p^{\prime2}}
      \left( 1 - \frac{m_{0}^{2}}{p^{\prime2}}
      \log\left[1+\frac{p^{\prime2}}{m_{0}^{2}}\right] \right)
      \right\} \right. \nonumber \\
      & \hphantom{= \frac{g^{2}C_{F}}{\left(4\pi\right)^{2}}[ }
      \left.  - \left( 1 - \alpha \right) \left( F_{0000} - \gamma_{E}
      + 1 \right) + Z + ZT - \frac{3}{2} + \mathcal{O}\left(a\right)
      \right] \,,
  \label{eq:sig-1-fin-1}
  \\
  \Sigma_{2}\left(p'\right) & =
  \frac{g^{2}C_{F}}{\left(4\pi\right)^{2}} \left[
    \left(4-\alpha\right) \left\{ \log\left[ a^{2} \left( m_{0}^{2} +
      p^{\prime2} \right) \right] - \left( 1 -
    \frac{m_{0}^{2}}{p^{\prime2}}
    \log\left[1+\frac{p^{\prime2}}{m_{0}^{2}}\right] \right) \right\}
    \vphantom{\frac{3}{2}} \right. \nonumber \\
    & \hphantom{= \frac{g^{2}C_{F}}{\left(4\pi\right)^{2}}[ } \left.
      \vphantom{\frac{3}{2}} - \left(4-\alpha\right) \left(
      F_{0000} - \gamma_{E} + 1 \right) + ZM + \mathcal{O}\left(a\right)
      \right] \,.
        \label{eq:sig-z-2-2}
\end{align}
Here, $Z+ZT=0.7737683(12)$, $ZM=13.242431(11)$, and $F_{0000} -
\gamma_{E} + 1 =4.7920095689746(13)$ are numerical results for the
Feynman diagrams obtained using the VEGAS algorithm
\cite{Lepage:1977sw}.

The inverse quark propagator $S^{-1}$ can now be expressed as
\begin{align}
  S^{-1} &= S_{0}^{-1} - \Sigma\left(p'\right) = -i
  \left(1-\Sigma_{1}\right) p'_{\mu}
  \overline{\overline{\left(\gamma_{\mu}\otimes\mathbf{1}\right)}}_{AB}
  + \left(1-\Sigma_{2}\right) m_{0}
  \overline{\overline{\left(\mathbf{1}\otimes\mathbf{1}\right)}}_{AB}
  \,.
  \label{eq:s_inv}
\end{align}
As usual, this form holds to all orders in perturbation theory, although here we use
only the one loop form of $\Sigma_{1,2}$.
%
%

\section{Results for $Z_q$ in the RI-MOM scheme}
\label{sec:def-zq}

Using the Ward identity following from the conservation of the vector current
(which holds also with staggered fermions) one can
derive the following identity for $Z_q$ in the RI-MOM scheme~\cite{Martinelli:1994ty}:
\begin{align}
  Z_{q} &= \left. \frac{1}{48} \tr \left( \frac{i}{4} \sum_{\rho}
  \overline{\overline{\left(\gamma_{\rho}\otimes\mathbf{1}\right)}}
  \frac{\partial}{\partial p'_{\rho}} S^{-1}\left(p'\right) \right)
  \right|_{p^{\prime 2} = \mu^{2}} \,.
  \label{eq:z_q:wi}
\end{align}
Substituting Eq.~\eqref{eq:s_inv} into this result we find
\begin{equation}
  Z_q = \left. \left( 1-\Sigma_{1}\left(p'\right) - \frac{1}{4}
  \sum_{\rho} p'_{\rho} \frac{\partial \Sigma_{1}\left(p'\right)}{\partial p'_{\rho}}
  \right) \right|_{p^{\prime2}=\mu^{2}} \,.
    \label{eq:ri-full-zq-3}
\end{equation}
Similarly, implementing the RI$'$-MOM scheme of Ref.~\cite{Martinelli:1994ty} 
(as has been done with staggered fermions in Ref.~\cite{Lytle:2013qoa}) we find
\begin{equation}
  Z_q' = \left. \vphantom{\frac{1}{12}}
  \left(1-\Sigma_{1}\left(p'\right)\right)
  \right|_{p^{\prime2}=\mu^{2}} \,.
  \label{eq:zq-rip-self-1}
\end{equation}
The result differs only in the absence of the last term in Eq.~\eqref{eq:ri-full-zq-3}.

Combining results for $\Sigma_1$ in Eq.~\eqref{eq:sig-1-fin-1} in the
chiral limit ($m_0=0$) with the master formulae
Eqs.~\eqref{eq:ri-full-zq-3} and \eqref{eq:zq-rip-self-1}
we obtain the one-loop results for $Z_q$ and $Z'_q$:
\begin{align}
    Z_q(\mu) & = 1 - \frac{g^{2}C_{F}}{\left(4\pi\right)^{2}} \left[ Z + ZT
      - \frac{3}{2} + \left(1-\alpha\right) \left( \log\left[ a^{2}
        \mu^2 \right] + \frac{1}{2} - X \right) +
        \mathcal{O}\left(a\right) \right]\,,
  \label{eq:zq-ri-fin-ml-1}
\\
  Z_q'(\mu) & = 1 - \frac{g^{2}C_{F}}{\left(4\pi\right)^{2}} \left[ Z + ZT
      - \frac{3}{2} + \left(1-\alpha\right) \left(\log\left[ a^{2}
        \mu^2 \right] - X \right) +
        \mathcal{O}\left(a\right) \right] \,,
  \label{eq:zq-rip-fin-ml-1}
\end{align}
where $X \equiv F_{0000} - \gamma_{E} + 1$.

To evaluate $Z_q$ and $Z'_q$ for scales $\mu$ far from $1/a$, we use
the horizontal matching method of Refs.~\cite{ Gupta:1996yt,
  Bae:2010ki}.  We first set $\mu = 1/a$ in the one-loop results, and
then evolve $Z_q$ from $\mu = 1/a$ to the final scale $\mu_0$ with
four-loop running, using the results from Ref.~\cite{
  Chetyrkin:1999pq}.
Typically we use $\mu_0 = 2\GeV \text{ or } 3 \GeV$.
The result is
\begin{align}
  Z_{q}(\mu_0) &= \frac{c(
    \mu_0)}{c(\mu)}
  Z_{q}\left(\mu=1/a\right) \,,
  \qquad
  Z_{q}'(\mu_0) = \frac{c'(
    \mu_0)}{c'(\mu)}
  Z_{q}'(\mu=1/a) \,, \label{eq:c-x}
\end{align}
where the prefactors $c(\mu_0)/c(\mu)$ and $c'(\mu_0)/c'(\mu)$ 
are the four-loop RG running factors in the RI-MOM and RI$'$-MOM schemes, respectively.

When using these formulae to give numerical values for $Z_q$ and $Z'_q$ 
we estimate the systematic error due to truncating the perturbative series at one loop
by assuming an $\mathcal O(1)$ coefficient of the missing $\alpha_s^2$ term.
Specifically, following Ref.~\cite{Bae:2011ff}, we estimate the truncation error to be
\begin{equation}
  E_{\text{trunc.}}^{(\prime)} \approx Z_{q}^{(\prime)} \times
  \alpha_{s}^2(\mu=1/a) \,.
  \label{eq:sys-err-1}
\end{equation}
Specifically, we use $\alpha_{s}\left(\mu=1/a\right)$ in the $\MSbar$ scheme,
evaluated from the PDG value for $\alpha_s(M_Z)$ using four-loop running.

%
\section{Numerical results for $Z_q$}
\label{sec:num-res}

%

%
%

\begin{figure}[t!]
  \begin{subfigure}{0.45\linewidth}
    \vspace*{-7mm}
    \centering
    \includegraphics[width=\textwidth]{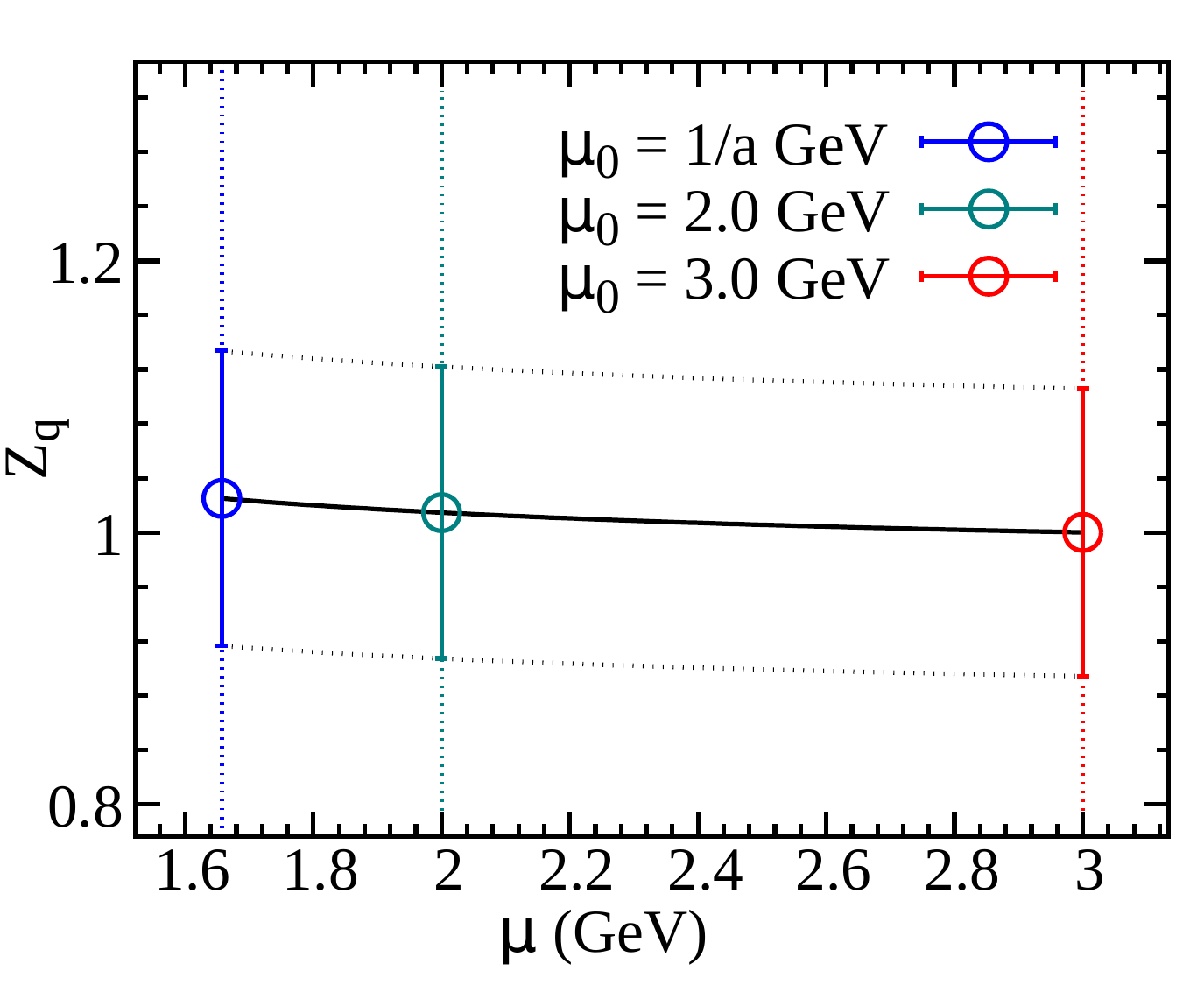}
    \caption{$Z_q$}
    \label{fig:Z_q}
  \end{subfigure}
  \hfill
  \begin{subfigure}{0.45\linewidth}
    \vspace*{-7mm}
    \centering
    \includegraphics[width=\textwidth]{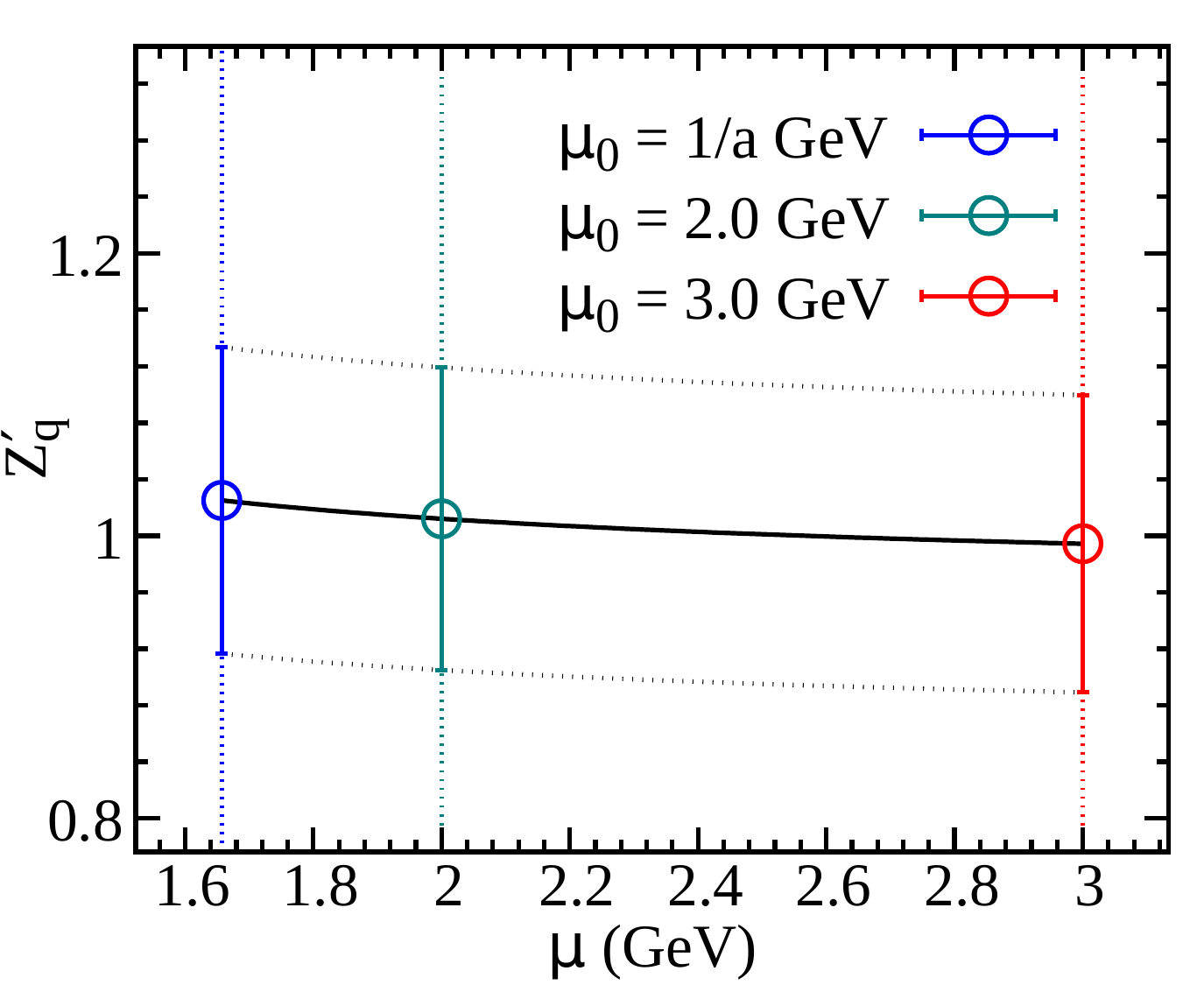}
    \caption{$Z_q'$}
    \label{fig:Z_q'}
  \end{subfigure}
  \caption{ One loop perturbative results for $Z_q$ and $Z_q'$ in Landau gauge.}
  \label{fig:Z_q:final}
\end{figure} 

In Fig.~\ref{fig:Z_q:final}, we present results for  $Z_q$ and $Z'_q$, evaluated
in perturbation theory in Landau gauge, at three scales:
$\mu_0=1/a$ (blue circles),
$\mu_0=2\GeV$  (green circles) and 
$\mu_0=3\GeV$ (red circles),
using  Eq.~\eqref{eq:c-x}.
In Table \ref{tab:Zq:fin}\subref{tab:Zq:fin-one-loop}, we give the corresponding numerical
values.

\begin{figure}[t!]
  \begin{subfigure}{0.45\linewidth}
    \centering
    \includegraphics[width=\textwidth]{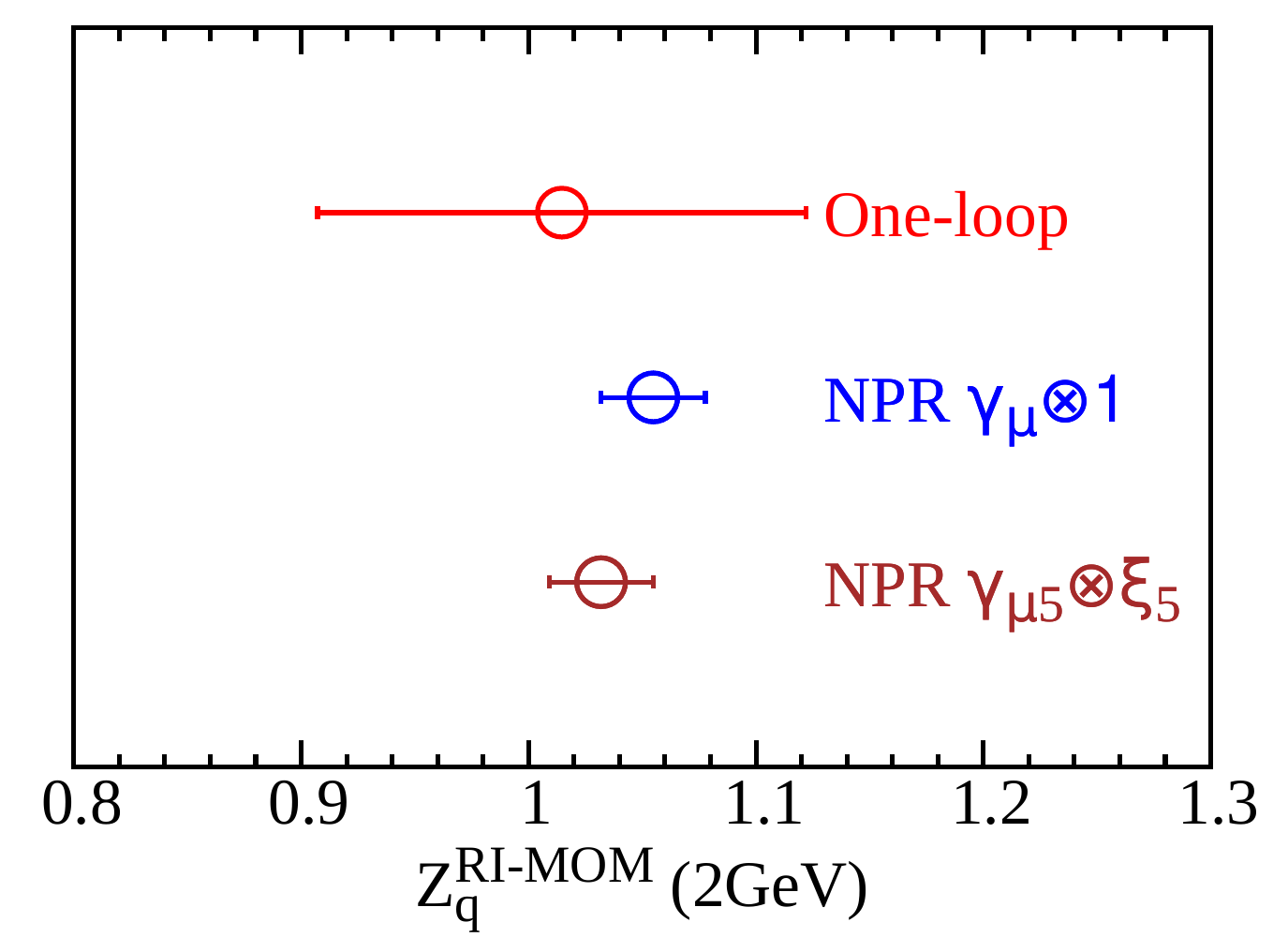}
    \caption{$\mu_0=2 \GeV$}
    \label{fig:cmp-2GeV}
  \end{subfigure}
  \hfill
  \begin{subfigure}{0.45\linewidth}
    \centering
    \includegraphics[width=\textwidth]{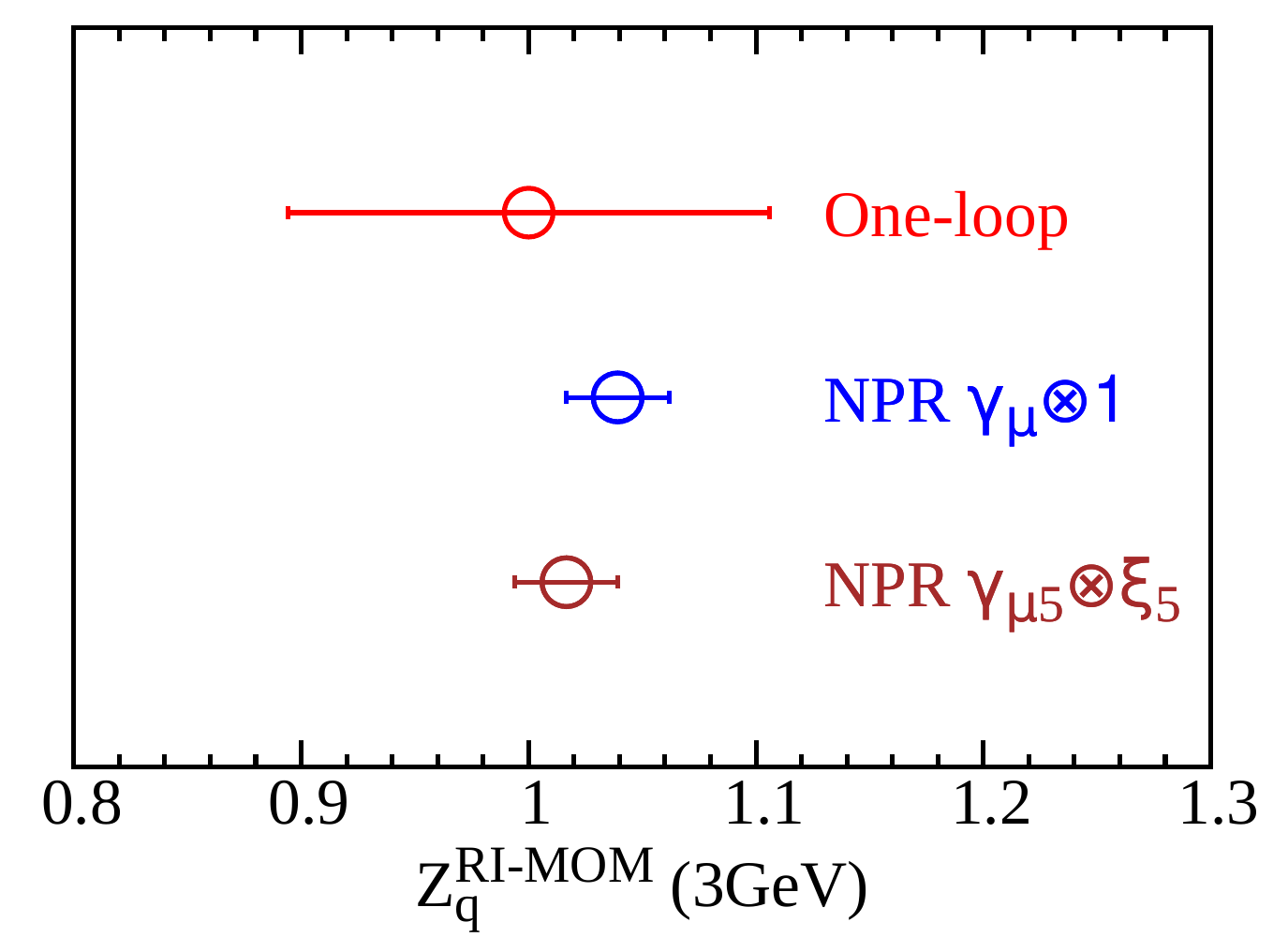}
    \caption{$\mu_0=3 \GeV$}
    \label{fig:cmp-3GeV}
  \end{subfigure}
  \caption{Comparison of results for $Z_{q}(\mu_0)$ between one-loop
    perturbation theory and NPR.}
  \label{fig:cmp}
\end{figure}

We now compare the results with those obtained using nonperturbatively.
We have results for $Z_q$ in the RI-MOM scheme using NPR with
HYP-smeared staggered fermions
on the MILC asqtad coarse lattice ensemble 2064f21b676m010m050, for
which $1/a = 1.657(2) \GeV$~\cite{Kim:2012ng,Kim:2013bta}.
We present these results in Table
\ref{tab:Zq:fin}\subref{tab:Zq:fin-npr}.
Here, NPR $\gamma_\mu \otimes \mathbf{1}$ ($\gamma_{\mu5}\otimes
\xi_5$) indicates results for $Z_q$ obtained using the conserved
vector (axial) current.
The errors are, respectively, statistical and systematic.
%
%
\begin{table}[h!]
  \begin{subtable}{0.54\linewidth}
    \renewcommand{\arraystretch}{1.2}
    \resizebox{1.0\linewidth}{!}{
  \begin{tabular}{ @{\quad} l @{\quad} | l  l | l  l }
    \hline\hline
    & \multicolumn{2}{c|}{Landau Gauge} & \multicolumn{2}{c}{Feynman Gauge}
    \\ 
    & $\mu_{0}=2\GeV$ & $\mu_{0}=3\GeV$ & $\mu_{0}=2\GeV$ & $\mu_{0}=3\GeV$
    \\ \hline
    $Z_{q}$  & 1.02(11) & 1.00(11) & 1.14(12) & 1.09(12)
    \\ \hline
    $Z_{q}'$ & 1.01(11) & 0.99(11) & 1.15(12) & 1.09(12)
    \\ \hline\hline
  \end{tabular}
    } 
  \caption{One-loop perturbation theory}
  \label{tab:Zq:fin-one-loop}
  \end{subtable} 
  \hfill
    \begin{subtable}{0.44\linewidth}
    \renewcommand{\arraystretch}{1.2}
    \resizebox{1.0\linewidth}{!}{
  \begin{tabular}{ @{\quad} c @{\quad} | l  l }
    \hline\hline
    \multirow{2}{*}{$Z_{q}$} & \multicolumn{2}{c}{NPR (Landau gauge)}
    \\ 
    & $\mu_{0}=2\GeV$ & $\mu_{0}=3\GeV$
    \\ \hline
    $\gamma_\mu \otimes \mathbf{1}$  & 1.0548(59)(229) & 1.0392(58)(226)
    \\ \hline
    $\gamma_{\mu 5} \otimes \xi_{5}$  & 1.0319(61)(229) & 1.0166(61)(226)
    \\ \hline\hline
  \end{tabular}
    } 
  \caption{NPR}
  \label{tab:Zq:fin-npr}
  \end{subtable} 
  \caption{Results on $Z_{q}$ and $Z_{q}'$ at $\mu_{0}=2\,\GeV$ and
    $3\,\GeV$.}
  \label{tab:Zq:fin}  
\end{table}
A graphical comparison is shown in Fig.~\ref{fig:cmp}.
We find that all the results are consistent within the quoted
uncertainties, with the results from NPR being significantly more accurate.
%

%

\acknowledgments
%
W.~Lee would like to acknowledge the support from the KISTI
supercomputing center through the strategic support program for the
supercomputing application research (No. KSC-2016-C3-0072).
The research of W.~Lee is supported by the Creative Research
Initiatives Program (No. 2017013332) of the NRF grant funded by the
Korean government (MEST).
The work of S.~Sharpe was supported in part by the US DOE grants
no.~DE-FG02-96ER40956 and DE-SC0011637.
Computations were carried out in part on the DAVID GPU clusters at
Seoul National University.

\bibliography{ref}

\end{document}